\documentclass[pra,twocolumn,graphicx,amssymb,floatfix]{revtex4}
\usepackage{graphicx}
\begin{document}

\title{Counterfactuality of ``counterfactual'' communication }

\author{L. Vaidman}
\affiliation{ Raymond and Beverly Sackler School of Physics and Astronomy\\
 Tel-Aviv University, Tel-Aviv 69978, Israel}

\begin{abstract}
 The counterfactuality of  the recently proposed protocols for direct quantum communication is analyzed. It is argued that the protocols can be counterfactual only for one value of the transmitted bit. The protocols achieve a reduced probability of detection of the particle in the transmission channel by increasing the number of paths in the channel. However, this probability   is not  lower than the probability of detecting a  particle actually passing through such a multi-path channel, which was found to be surprisingly small. The relation between security and counterfactuality of the protocols is discussed. An analysis of counterfactuality of the protocols in the framework of the Bohmian interpretation is performed.
     \end{abstract}
\maketitle

\section{Introduction}

Penrose \cite{Penrose} coined the term ``counterfactuals'' for describing quantum interaction-free measurements (IFM) \cite{IFM}.
\begin{quote}
Counterfactuals are things that might have happened, although they did not in fact happen.

\hfill {Penrose 1994}
\end{quote}
He argued that in the IFM, an object is found because it might have absorbed a photon, although actually it did not.
The idea of the IFM has been applied to ``counterfactual computation'' \cite{Joz}, a setup in which one particular outcome of a computation becomes known in spite of the fact that the computer did not run the algorithm.  Noh \cite{Noh} created counterfactual cryptography, a method for  secret key distribution using events in which the particle  was not  present in the transmission channel. Noh used a random choice of orthogonal input states (like in \cite{GV}) in contrast with the non-orthogonal states of BB84 cryptographic protocol \cite{BB84}.

 It was argued \cite{Ho06}, that a modification of the counterfactual computation proposed above, which includes quantum Zeno effect,  can achieve  counterfactuality for {\it all} outcomes  of the computation. Recently, this idea has been used for performing ``counterfactual communication'' \cite{Salih}, which supposedly allowed to send information from Bob to Alice without transferring any particle between them. The transmission happens in a counterfactual way: the mere possibility of transmitting the particle allows transmitting  the value of the bit.

I find all these results very paradoxical since they contradict  physical intuition of causality: information is usually  transmitted  continuously in space.  I argued \cite{PSA} that to resolve the paradoxical feature of the IFM  one has to adopt the many-worlds interpretation (MWI) of quantum mechanics \cite{Eve,myMWI}, in which the particle touches the object in a parallel world restoring causality at least within the complete physical universe which includes all the worlds. However, I believe that a protocol which can transmit both values of a bit without any particle present in the transmission channel is   impossible,  irrespectively of the interpretation of quantum mechanics one adopts. I have expressed this opinion already \cite{Va07,MyCom}, but more protocols were suggested \cite{Li} and  the controversy remains open \cite{LiCom,LiComRep,SalihReply}. The clarification of these conceptual issues is particularly important due to the recent increasing interest in the applications of counterfactual protocols \cite{CFInt1,CFInt2,CFInt3,CFInt4,CFInt6,CFInt7,CFInt8,CFInt9,CFInt10}. Here I  discuss this issue in more detail and try to resolve the controversy.

The plan of the paper is as follows. In Section II I introduce the general setup of quantum communication protocols. In Sections III and IV I describe two recent protocols which are claimed to be counterfactual. In Section V I analyze various possible definitions of counterfactuality and define my preferred criterion which is based on the magnitude of the trace left in the transmission channel.  In Section VI  I calculate the trace left by a single particle present in the channel, i.e. the trace of a non-counterfactual communication protocol. In Sections VII-IX I  show that the trace in the protocols claimed to be ``counterfactual'' is not less than the trace in a non-counterfactual protocol. In Section X I analyze the security of ``counterfactual'' protocols against an eavesdropper.   Section XI is devoted to counterfactuality in the framework of the Bohmian interpretation. I summarize the results in Section XII.

\section{Communication with quantum particles}\label{ComQP}

There is a surprisingly low bound on the number of bits which can be sent using 1 qubit: The Holevo bound of 1, when the qubit is not entangled \cite{Hol}, and  2, when entanglement is allowed \cite{supcod}. This is when we transmit one particle with an internal structure of a qubit such as a polarization state of a photon. In this paper I analyze protocols in which  the particle does not have an internal structure: the information is encoded in the presence or absence of the particle.

Let Alice and Bob be on the two separate sides of a region, see Fig.~1. Bob has a mirror on his side which causes particles sent by Alice to be bounced back to her.
For a bit value of 1, Bob places a shutter which absorbs Alice's particles, while for  0, the shutter is absent.
\begin{figure}
\begin{center}
 \includegraphics[width=7.2cm]{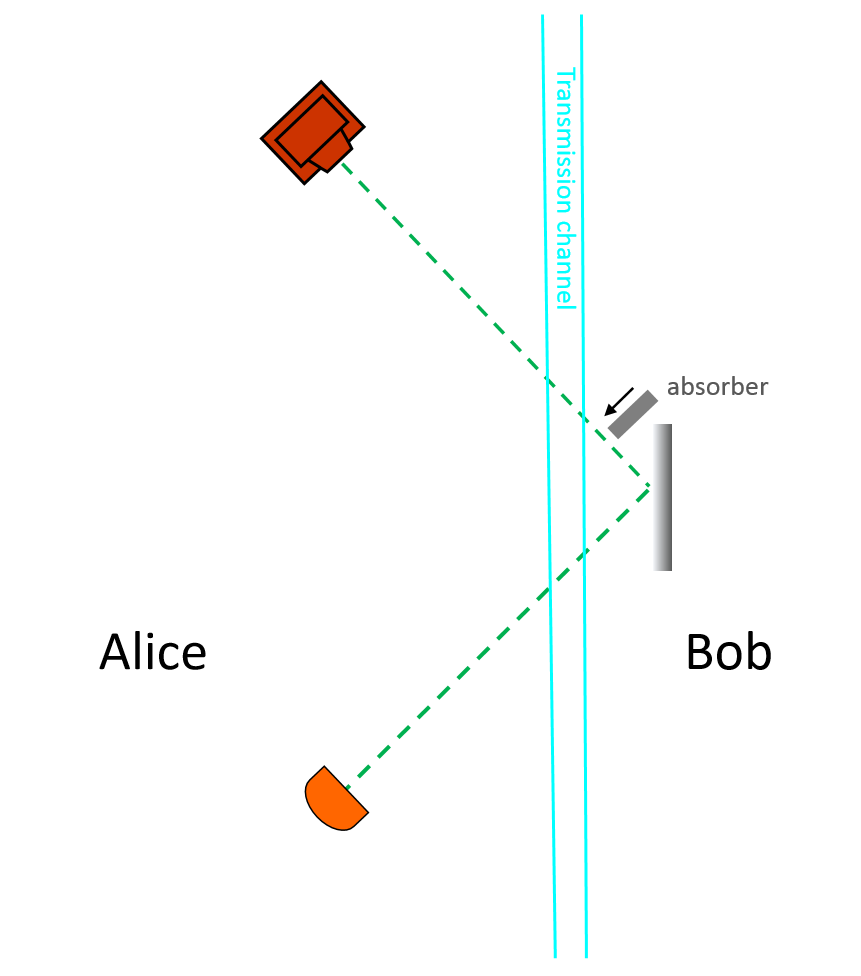}\end{center}
\caption{   Simple communication with a quantum particle. Alice sends a particle to Bob and knows the bit chosen by Bob through observation whether or not the particle comes back to her. }
\label{fig:1}
\end{figure}

Quantum mechanics, via IFM,  allows, at least sometimes, to transmit the bit 1 in a counterfactual manner, i.e., without having any  particle in the transmission channel, see Fig.~2. Alice arranges a Mach-Zehnder interferometer (MZI) tuned to destructive interference in one of the ports, say $D_1$, such that one arm of the interferometer crosses the place where Bob's shutter might be. Detection of the particle in the dark port of the interferometer tells her with certainty that the bit is equal to 1 (the shutter is present).

\begin{figure}
\begin{center}
 \includegraphics[width=7.2cm]{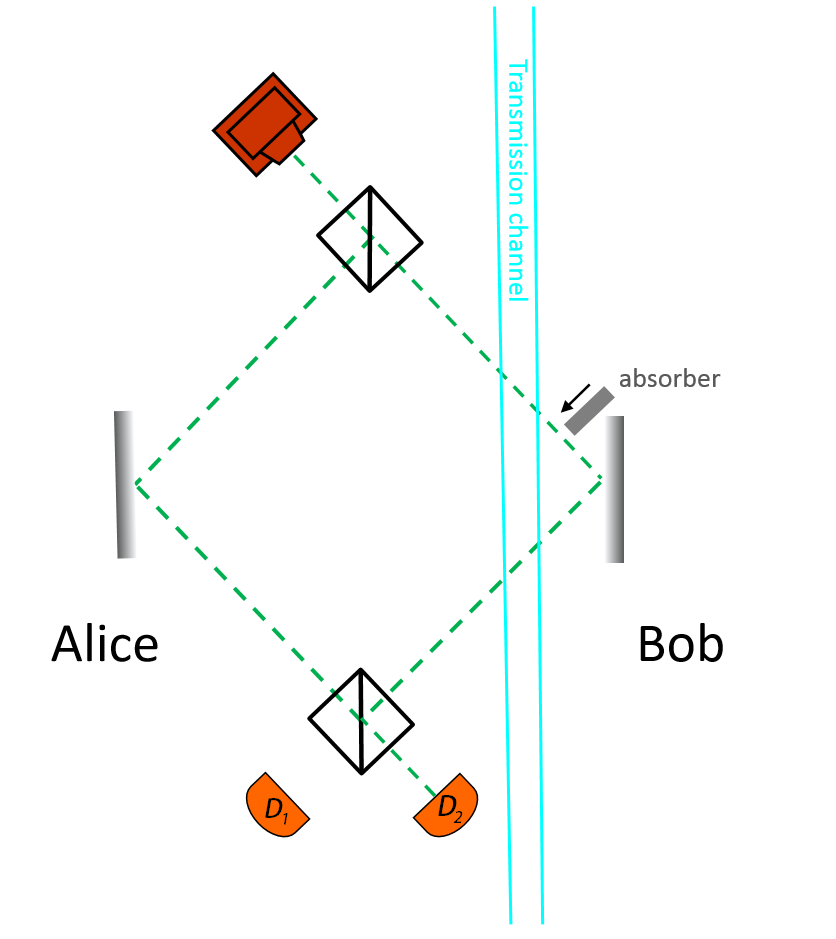}\end{center}
\caption{ A single bit value communication with IFM. The interferometer is tuned in such a way that  detector $D_1$ never clicks if the paths are free. Alice  knows that  bit 1 has been chosen (Bob blocked the path) when she observes the click in $D_1$. }
\label{fig:2}
\end{figure}

The simplest argument that in this case the particle was not present in the transmission channel is: ``If the particle were in the transmission channel, it could not be detected by Alice''. In my view, this argument cannot be used for claims about quantum particles \cite{past}. I, instead, suggest to rely on the fact that the particle does not leave any trace in the transmission channel.

Note that counterfactual transmission of just one bit value  can be achieved using a classical particle \cite{Gisin}. Alice and Bob agree in advance that at a particular time, for a particular value of a bit, Bob sends  the particle to Alice, while for the other bit value, he sends nothing. Note, however, that this classical protocol cannot achieve the task of the quantum IFM. In the IFM, Alice learns about the shutter in Bob's place without prior agreement with Bob and without Bob knowing that she acquired this information.

In the IFM shown in  Fig.~2, Alice does not obtain a definite information about the classical bit 0. Without the shutter, the click in the bright port happens with certainty, but  it might happen  (with probability 25\%) with the shutter too. This protocol is also not the most efficient method for communication of the bit 1. When the particle is detected by a bright port (probability 25\%) we get no decisive information, and in half of the cases the particle is lost (then we get the information that the bit is 1 but not in a counterfactual manner).

The quantum method can be modified to be a  reliable transmission of both values of the  bit. To this end, instead of the shutter, Bob inserts a half-wave plate (HWP), see Fig.~3. This transforms the dark port to bright port and vice versa. However, half of the wave always passes through the communication channel, so one cannot argue   that this is a counterfactual communication.
\begin{figure}
\begin{center}
 \includegraphics[width=8.2cm]{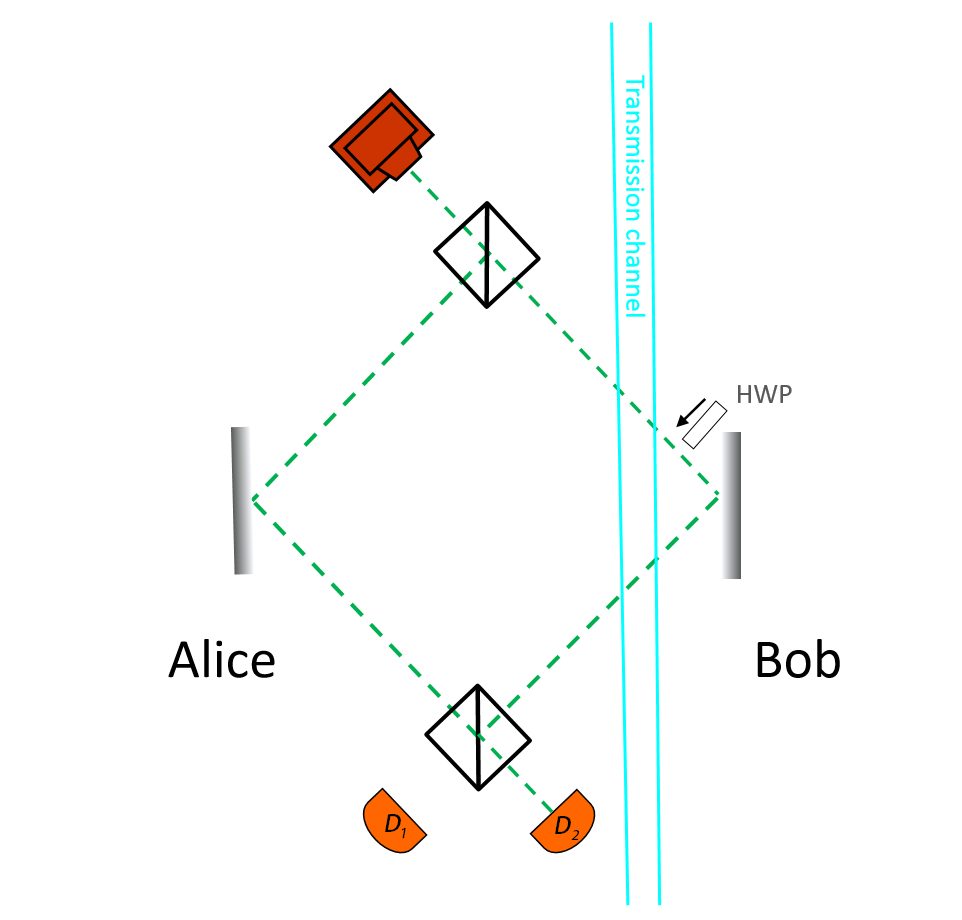}\end{center}
\caption{ Communication with MZI and HWP. The interferometer is tuned to destructive interference towards  $D_1$. Bob communicates the bit to Alice by changing the destructive interference to detector $D_2$ by inserting the HWP in the right path of the particle. }
\label{fig:3}
\end{figure}

\begin{figure}[b]
\begin{center}
 \includegraphics[width=7.2cm]{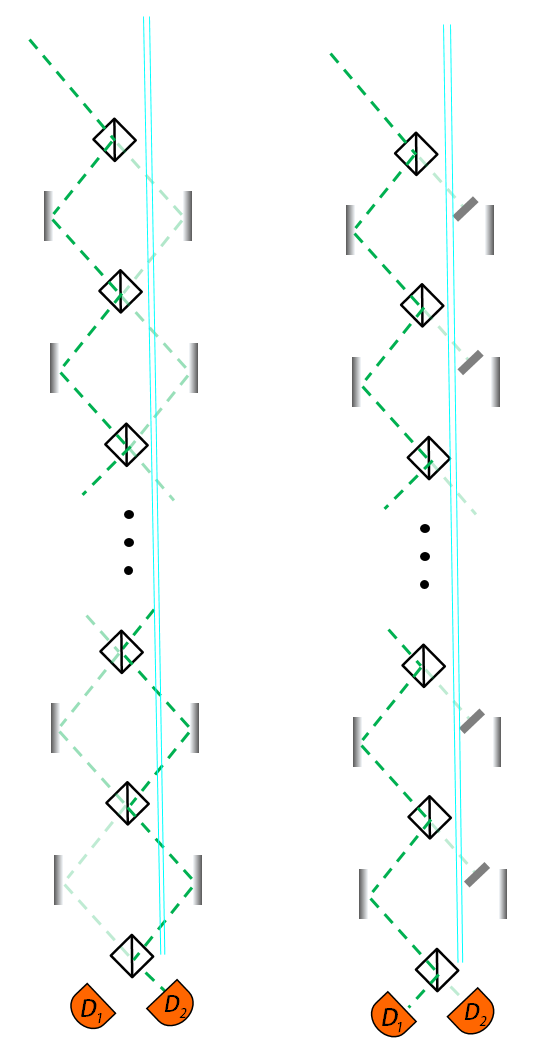}\end{center}
\caption{ Efficient communication using IFM and quantum Zeno effect. a). The chain of interferometers wih highly reflective beam splitters is tuned such that the particle has destructive interference towards $D_1$. b). Bob blocks the interferometers and then, due to Zeno effect, the particle reaches detector $D_1$ with probability close to 1. }
\label{fig:4}
\end{figure}

Consider next what happens when we combine the quantum Zeno effect with the IFM \cite{ZenoIFM}. It allows to perform a counterfactual transmission of bit 1  with probability arbitrary close to 1. The device consists of  a  chain of $N$ interferometers  with identical beam splitters with  small transmittance ${T_1=\sin^2 \alpha}$, see Fig.~4a. Each one of the beam splitters  in the chain performs  the following unitary evolution of the state of the particle:
\begin{eqnarray}
\label{BS} \nonumber
|L\rangle &\rightarrow & \cos   \alpha   |L\rangle + \sin \alpha |R\rangle,\\
 |R\rangle &\rightarrow & -\sin \alpha |L\rangle + \cos \alpha
 |R\rangle .
\end{eqnarray}
A straightforward calculation shows that $n$ beam splitters  perform   the following   evolution of the wave packets of the particle entering the chain:
\begin{eqnarray}\label{n-steps}
\label{BS} \nonumber
|L\rangle &\rightarrow & \cos   n\alpha   ~|L\rangle + \sin n\alpha ~|R\rangle,\\
 ~|R\rangle &\rightarrow & -\sin n\alpha |L\rangle +  \cos n\alpha
 ~|R\rangle .
 \end{eqnarray}
For the particular choice of the transmittance parameter,  $\alpha = \frac {\pi}{2N}$, after passing all $N$ beam splitters, the wave packet of the particle moves  from one side to the other: $|L\rangle  \rightarrow |R\rangle$, ${|R\rangle  \rightarrow -|L\rangle}$.

 The Zeno effect takes place when the right arms of the interferometers are blocked, Fig.~4b.  The state remains $|L\rangle$  with   probability close to 1  when $N$ is large, $ \cos^{2N} \frac{\pi}{2N} \simeq  1-\frac {\pi^2}{4N} $.

In summary, for bit  0 Bob does nothing  and Alice gets the click with certainty  at the right  port in the detector $D_2$.   For bit  1 Bob blocks the interferometers and Alice gets the click with a very high probability  in the detector $D_1$.

\section{``Direct Counterfactual Quantum Communication"}\label{DirectCF}

\begin{figure}
\begin{center}
 \includegraphics[width=6.2cm]{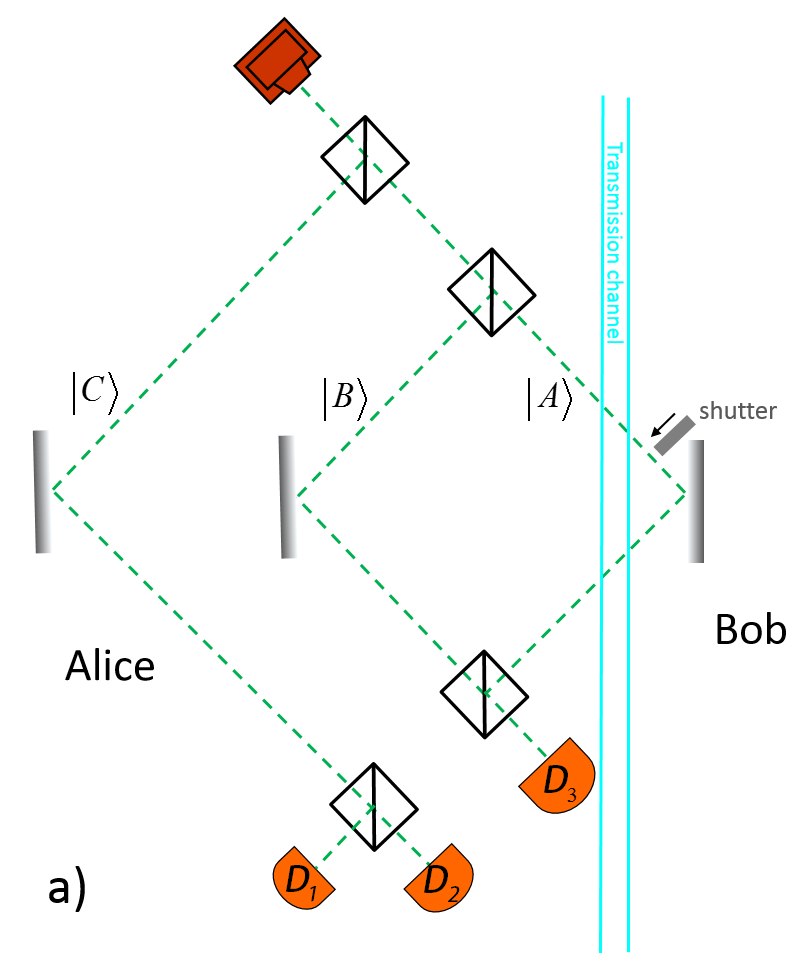} \includegraphics[width=6.2cm]{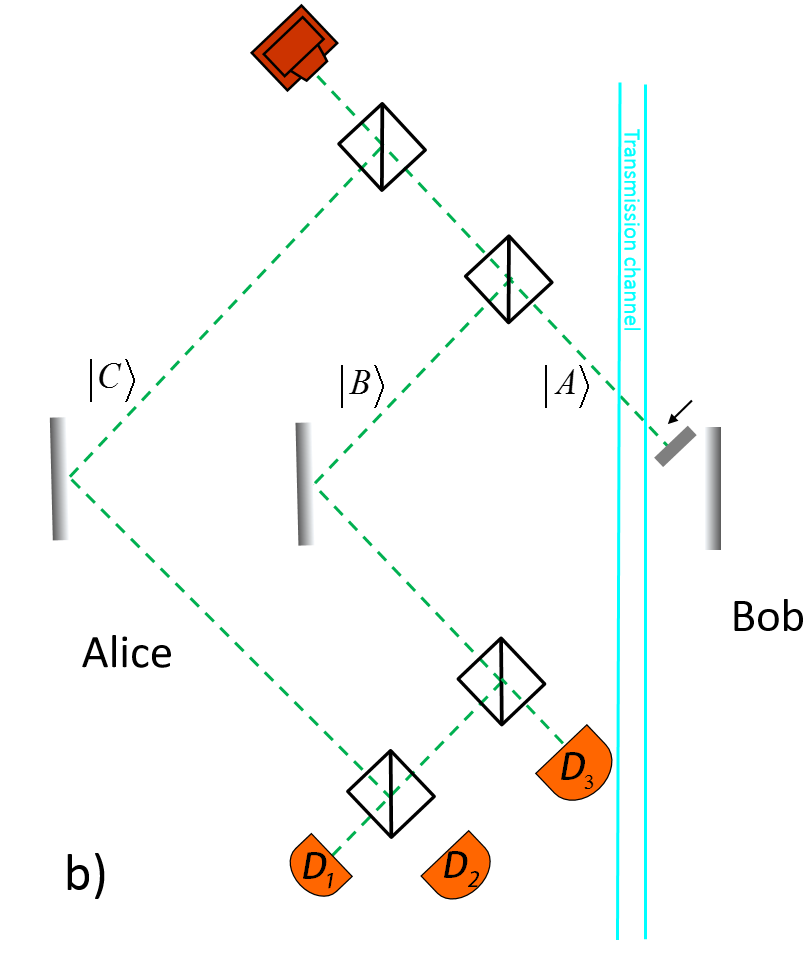}\end{center}
\caption{ Communication with nested MZIs. a). The inner interferometer is tuned such that the particle cannot pass through the right arm of the external interferometer. b). There is a destructive interference towards $D_2$ when the right path of the inner interferometer is blocked. }
\label{fig:5}
\end{figure}

In this section I  describe the recent protocol by Salih et al. \cite{Salih} which followed the idea of counterfactual  computation  \cite{Ho06,Va07}.

Let us first consider a  MZI nested in  another MZI,  see Fig.~5. The inner interferometer is tuned for destructive interference toward the second beam splitter of the external interferometer, see Fig.~5a. The external interferometer is tuned for destructive interference towards $D_2$ when the lower path of the inner interferometer is blocked,  see Fig.~5b.  This configuration provides (sometimes) definite information about value  0 of the bit, namely the absence of the shutter. Indeed, click in $D_2$ is possible only if the shutter is not present. One can  naively argue that  Alice obtains this information in a counterfactual way, since the particle cannot pass through Bob's site and reach  $D_2$. However, as detailed  in Section \ref{Criteria}, the presence of a weak trace inside the inner interferometer contradicts it.

\begin{figure*}
\begin{center}
 \includegraphics[height=15.2cm]{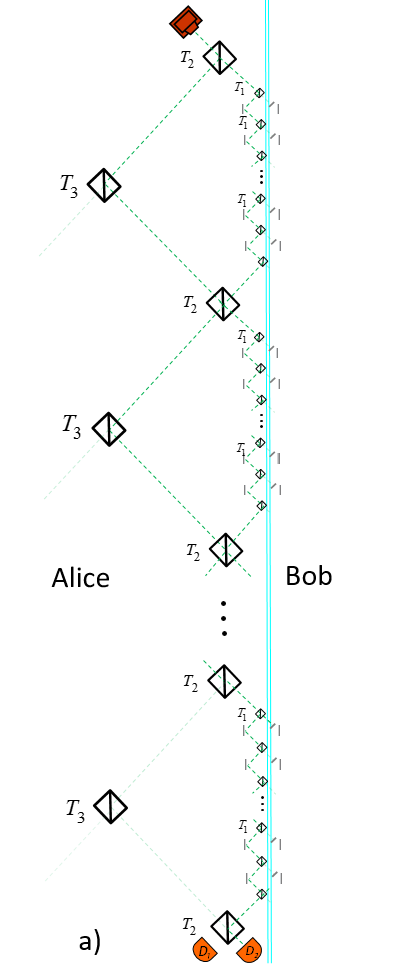}  \includegraphics[height=15.2cm]{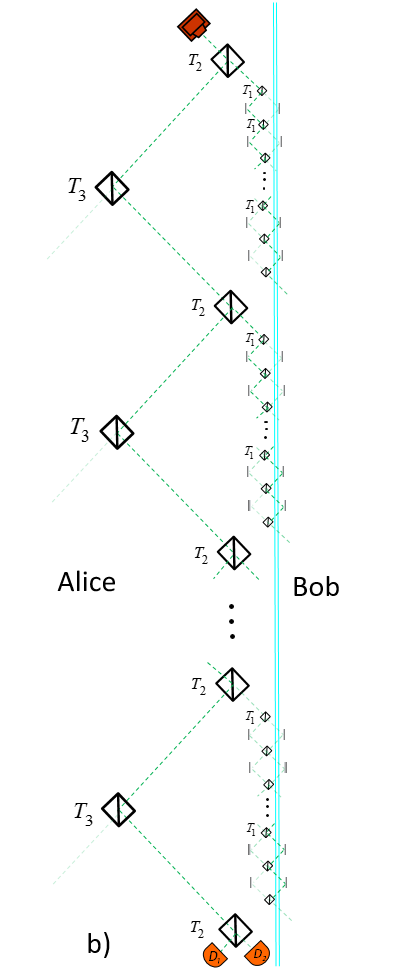} \end{center}
\caption{ ``Direct counterfactual quantum communication'' a). For bit 1 Bob blocks all inner interferometers. In this case detector $D_2$ clicks with probability close to 1, while $D_1$ cannot click.  b). For bit 0 Bob leaves all inner interferometers undisturbed. For large $N$,   $D_1$ clicks with probability close to 1, while the probability for a click in $D_2$ goes to zero.}
\label{fig:6}
\end{figure*}

 Salih et al. \cite{Salih} further argued that a scheme with numerous nested interferometers  leads to a protocol for transmitting both values of the bit in a counterfactual way with an efficiency which is arbitrarily close to 100\%.
In the protocol,    $M-1$   chains of the $N-1$ interferometers described in Fig.~4 are parts of  another  chain of interferometers with  $M$ beam splitters  having transmittance, $T_1=\sin^2 \frac {\pi}{2M}$. To simplify the analysis, I modify  Salih et al. protocol making it slightly less efficient, but still good according to their line of argument. I replace the mirrors of the external chain of interferometers by highly reflective beam splitters. Transmitted waves are lost, but the modification balances the losses in the inner chains when shutters are introduced, such that the states of particles which are not lost are still described by the same equation (\ref{n-steps}).

 The setup is described in Fig.~6. The external chain of interferometers has $M$ beam splitters with transmittance  $T_2=\sin^2 \frac{\pi}{2M}$ and the transmittance of the side beam splitters serving as mirrors is $T_3=1-\cos^{2N} \frac{\pi}{2N}$.

All right mirrors of the internal chains are in   Bob's territory. He knows that Alice sends a particle at a particular time at the top of the external chain in the  state  $|L\rangle $. For communicating bit 1, Bob blocks all inner interferometers, see Fig.~6a. Then, after $m$ beam splitters of the external interferometer, the {\it normalized} quantum state is
\begin{equation}\label{j-step}
 |\Psi^{(1)}_m\rangle =  \cos^{(m-1)N} \frac{\pi}{2N}\left (  \cos   \frac{m\pi}{2M}   |L\rangle + \sin \frac{m\pi}{2M} |R\rangle \right )+...~,
\end{equation}
and after all $M $ beam splitters the state is
\begin{equation}\label{M-step}
 |\Psi^{(1)}_M\rangle =    \cos^{(M-1)N} \frac{\pi}{2N} |R\rangle +...~.
\end{equation}
In both equations ``...'' signify states orthogonal to the term which is shown.  If $1 \ll M \ll N$, then the norm of the leading term in (\ref{M-step}) is close to 1:
 \begin{equation}\label{M-stepNORM}
 \cos^{2(M-1)N} \frac{\pi}{2N}  \simeq 1- \frac{\pi^2 M}{4N}.
\end{equation}
Thus, in the limit of large $N$, Bob's choice of bit 1  leads to a click of Alice's detector $D_2$. Since the state $|\Psi^{(1)}_M\rangle $ is orthogonal to the state  $|L\rangle$ at the output port of the interferometer, there is zero probability for a click of $D_1$. The particle can be lost in Alice's or Bob's territories, and then none of the detectors click, but  the probability of such a case vanishes for large $N$.

If Bob wants to communicate the bit 0, instead,  he does nothing,  Fig.~6b. Then, every wave packet entering any of the inner chains of the interferometers follows evolution (\ref{n-steps}) inside this chain and  does not come back to the external interferometer. At the output of the  interferometer, the   normalized  quantum state is
\begin{equation}\label{M0-step}
 |\Psi^{(0)}_M\rangle =   \cos^{(M-1)N} \frac{\pi}{2N} \cos^{M} \frac{\pi}{2M}  |L\rangle  +...
\end{equation}
Under the condition $1\ll M \ll N$,  the norm of the leading term in (\ref{M0-step}) is also close to 1:
 \begin{equation}\label{M-stepNORM}
 \cos^{2(M-1)N} \frac{\pi}{2N} \cos^{2M} \frac{\pi}{2M}   \simeq 1- \frac{\pi^2}{4}\left(\frac{M}{ N} +\frac{1}{M}\right).
\end{equation}
Therefore, the detection of  the particle in the left port by $D_1$ tells Alice that Bob sent bit 1.

Note,  that the probability for a failure might become large if the condition $1\ll M \ll N$ is not fulfilled. The particle might be lost or detected  by $D_2$. However, if the  condition holds, the probability for a failure is vanishingly small. The probability for loosing the particle and getting no result is of order   $ \frac{\pi^2 M}{4N}$ for bit 1 and  $\frac{\pi^2}{4}\left(\frac{M}{ N} +\frac{1}{M}\right)$ for bit 0.

The click of $D_2$ tells  Alice with certainty that the bit is 1, and the click of $D_1$ tells her that the bit is 0 with only a very small probability for an error: about $\frac{\pi^2}{4M^2}$. This is a good direct communication protocol.

\section{``Direct quantum communication with almost invisible photons''}\label{DirectnonZeno}

As in the simple example in Sec. \ref{ComQP} (Fig.~3), using HWPs instead of absorbing shutters leads to a communication protocol which is theoretically free of errors. Li et al. \cite{Li} suggest such a protocol and argue that it has  ``arbitrarily small probability of the existence of the particle in the transmission channel''.

The configuration is  similar to the experiment of Salih et al. \cite{Salih}: a chain of $M-1$ interferometers with inner chains of $N-1$ interferometers (but now $M,N$ have to be even numbers). Without absorbers, the evolution is unitary and the Zeno effect is not used in this protocol. There is no need to   modify the protocol by replacing   mirrors with beam splitters, because there are no losses to compensate.

The new protocol is different also in the transmittance of the beam splitters in the  inner chains. The parameter $\alpha$ is bigger by a factor of 2: $\alpha = \frac {\pi}{N}$. As a result, the chain (without the HWPs) works as two consecutive inner chains of the protocol discussed in the previous section. The first half of the chain moves the wave packet to the right side and the second brings it back to the left.  From (\ref{n-steps}) we obtain the transformation of the wave packet in the inner chain of the interferometers   $|L\rangle  \rightarrow -|L\rangle$, see Fig.~7a.

\begin{figure}
\begin{center}
 \includegraphics[width=6.2cm]{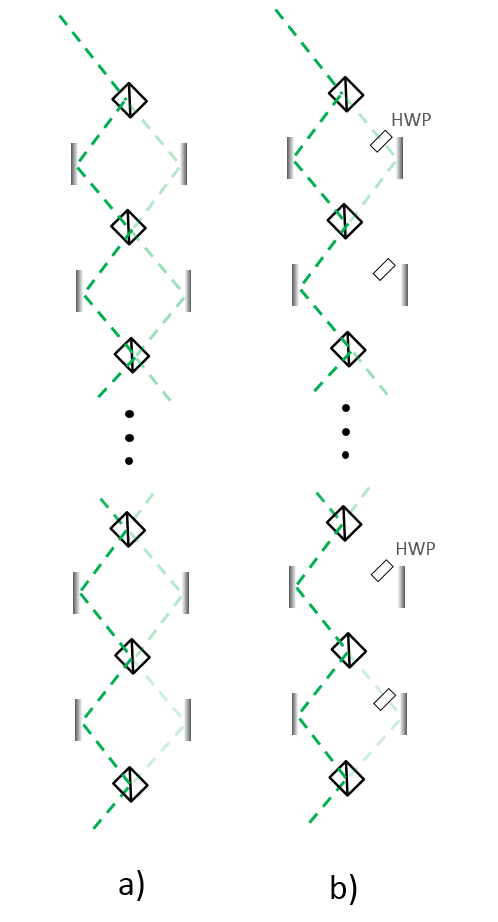}  \end{center}
\caption{   The chain of interferometers with highly reflective beam splitters manipulated by the HWPs.  a). The wave packet of the particle moves from left to right and then to left again but obtains the  phase $\pi$. b). HWPs ``undo'' the transformation on every second interferometer and  the wave packet ends up in the original state on the left without additional phase. }
\label{fig:7}
\end{figure}

When the HWPs are inserted in every interferometer of the inner chain, see Fig.~7b, they cause a $\pi$ phase change of every state $|R\rangle$ and, consequently, every second beam splitter reverses the operation of the previous one:
   \begin{equation}\label{2step}
 |L\rangle  \rightarrow \cos   \alpha   |L\rangle + \sin \alpha |R\rangle \rightarrow \cos   \alpha   |L\rangle - \sin \alpha |R\rangle \rightarrow |L\rangle .
\end{equation}
 Since every chain has an even number of beam splitters, the transformation of the wave packet in the inner chain  is  $|L\rangle  \rightarrow |L\rangle$.

\begin{figure*}
\begin{center}
 \includegraphics[height=15.2cm]{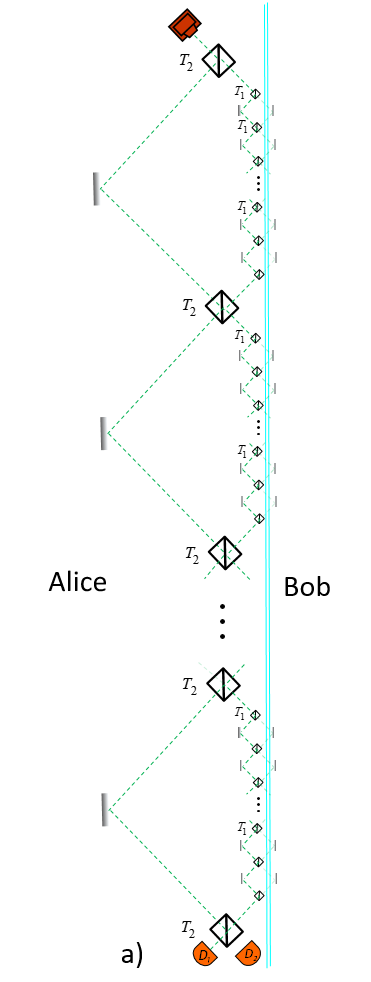}  \includegraphics[height=15.2cm]{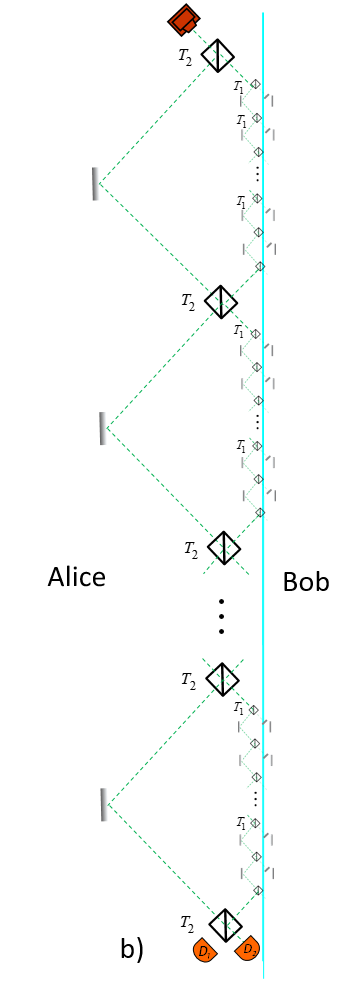} \end{center}
\caption{ ``Direct quantum communication with almost invisible photons'' a). The chains of the interferometers are tuned such that  $D_1$ clicks with probability  1. b). If Bob inserts HWPs in all inner interferometers, then $D_2$ clicks with probability  1. }
\label{fig:8}
\end{figure*}
The setup for sending bit 0 is described on Fig.~8a.  Bob leaves the inner interferometers untouched. Then, each inner chain of the interferometers changes the phase of the quantum state of the particle: $|L\rangle  \rightarrow - |L\rangle$. A state $|L\rangle$ of the inner interferometer is a state $|R\rangle$ of the external interferometer. Thus, the operation of the first external interferometer is
 \begin{equation}\label{2step}
 |L\rangle  \rightarrow  \cos   \alpha   |L\rangle + \sin \alpha |R\rangle \rightarrow \cos   \alpha   |L\rangle - \sin \alpha |R\rangle \rightarrow -  |L\rangle .
\end{equation}
 Since the number of beam splitters in the external chain is even, at the end of the process the particle is on the left side and it is detected by detector $D_1$ with certainty.

If Bob wants to communicate the bit 1, instead,  he inserts HWPs in all the interferometers of the inner chains    Fig.~8b.
Now, after every two beam splitters of the inner chain, the wave packet comes back to the left side without acquiring additional phase. Thus, every inner chain works as a mirror and the external chain of the interferometers moves the particle from left to right, to be detected with certainty by detector $D_2$. Alice knows with certainty the choice of Bob by observing which detector clicks. This is an ideal direct communication protocol: theoretically, when there are no losses, there is zero probability for  an error.

\section{Criteria for Counterfactuality }\label{Criteria}

The question I want to answer in this paper is: Can the  protocols of Sections \ref{DirectCF} and \ref{DirectnonZeno}
 be considered as {\it counterfactual} communication protocols?
 Let us consider the following three statements which try to capture the meaning of counterfactuality.

1) The probability of finding the particle in the transmission channel is zero or can be made arbitrarily small.

2) The particle did  not pass through the transmission channel.

3) The particle was not present in the transmission channel (or the probability of its existence
in the transmission channel can be made arbitrarily small).

 In the papers on counterfactual communication \cite{Salih,Li} all these statements  were  considered to be interchangeable, i.e. all are true and each one of them represents counterfactuality.   I argue that the situation is more subtle and clarification is needed.

Statement (1). A non-demolition measurement of the presence of the particle in the transmission channel disturbs completely the operation of  the communication protocols which are considered. When such a measurement is added to the protocol, Bob does not transmit information to Alice by his actions. So, the truth or falsehood  of this statement is not a decisive indication of the counterfactuality of the protocols.

However, since  there is a separate controversy about the validity of this statement for the two protocols under discussion, it should be clarified too. The papers on counterfactual communication claim that this is a correct statement while I \cite{MyCom} claim that  in these protocols the probability of finding the particle in the transmission channel is 1.

The source of this controversy is our different assumptions. The communication protocols  involve preparation and detection of the particle. I consider the probability of finding the particle in the transmission channel under the condition of the same final detection as in the  protocol without intermediate measurement. In this case, the probability of finding the particle is exactly 1, since had it it  not been found, the result of the final detection could not be that of the undisturbed protocol. On the other hand, without the condition on the result of the final detection, the probability to detect the particle in the transmission channel is vanishingly small. These are two correct statements about the probability of finding the particle in the intermediate measurement: the probability is 1 when both the proper preparation and the proper final detection are done, and it is vanishingly small  when only  the  preparation of the particle is assumed.  None of these statements shed much light on the issue of counterfactuality of the protocols {\it without} intermediate non-demolition measurements.

Statement (2). In contrast to such a claim for a classical particle, the meaning of this statement for a quantum particle is not well defined. For a classical particle, the operational meaning of (2) is  (1), but as discussed above, statement (1) is not helpful in the quantum case. In a double slit experiment with a screen, there is no good answer through which slit the particle passed and through which it did not pass. However, if the detector which finds the particle  is placed after one of the slits, and  the wave packet passing through the other slit does not reach the detector, then it is frequently declared, following Wheeler \cite{Whe}, that the particle did not pass through the second slit. (Note that this contradicts the textbook picture, attributed to von Neumann, according to which the wave passes through both slits and then collapses to the location of the detector.) If we adopt Wheeler's definition, then statement (2) is correct for the protocol of Section \ref{DirectCF}. The wave packets of the particle passing through the transmission channel do not reach the detector which detects the particle in this protocol. I, however, argued that we should not adopt Wheeler's definition for discussing the past of a quantum particle \cite{past}.

The concept of a quantum particle passing through a channel  has no clear meaning in the standard quantum mechanics in which particles do not have trajectories. It is rigorously defined in the framework of Bohmian mechanics \cite{Bohm52}.
 However, since the authors of papers on counterfactual communication never mentioned Bohmian mechanics,  it is not particularly relevant to the current controversy. Still, due to to its conceptual importance I, following the advice of a referee, will provide the analysis in the framework of the Bohmian interpretation, but only at the end of the paper, in Section \ref{Bohm}.

Statement (3). Apart from Bohmian mechanics,  quantum mechanics does not provide a rigorous meaning also for statement (3). Without a clear ontological definition  I suggest  to introduce an operational meaning. We cannot rely on an operational definition based on statement (1), since strong, even nondemolition, measurements change the setup we want to analyze. So, my proposal is to look at the {\it weak} trace the particle leaves.

All particles interact locally with the environment.  If the particle is present in a particular place, it leaves some trace there, and it does not leave any trace where it was not present. Therefore, we can run the protocol  we want to analyze, and then look at the trace left in the environment.  If in the transmission channel there is a non-zero trace, we will say that the protocol  is not counterfactual. The counterpart of (3),  a definition of counterfactual as a process without a local trace, is less robust, because, although very unlikely, it is possible that the particle changed the local environment via some local interaction, but then changed it back to its original state.

Since we are all along analyzing interference experiments, the trace left by the particles has to be small, as otherwise the interference is destroyed. When the trace is small, one may argue that it can be neglected. I, however, claim that it can be neglected only if it is small compare to the trace which a single particle with the same coupling being at the same place would leave.  Hence, the remaining  task is the comparison between the trace left in the transmission channel in the ``counterfactual'' communication protocols \cite{Salih,Li} and the trace left by a single particle passing through the channel. In the next section I will analyze the minimal trace left by a single particle being in a transmission channel and the comparison will be made in the following sections.

 \section{The trace left by a single particle }\label{presence}

In the framework of standard quantum mechanics there is no  rigorous way to decide if statement (3) holds, that is: whether or not the particle was present in the transmission channel. In  a two-slit experiment it is not clear whether the particle was in a particular slit. If a particle starts on one side of a plate with two slits and it is found later on the other side, we do not know if the particle was in the two slits together, or in one of them. However, we firmly believe that it cannot be that the particle was not present in both.

\begin{figure}
\begin{center}
 \includegraphics[width=7.2cm]{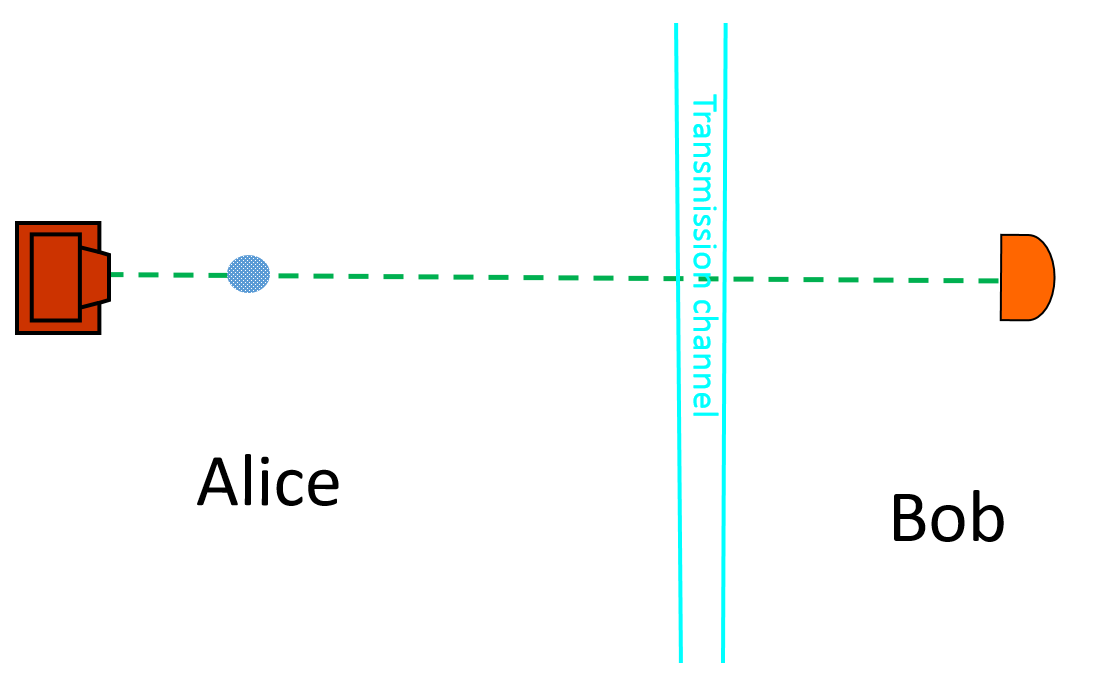}  \end{center}
\caption{ A single particle in a single localized wave packet passes from Alice to Bob  in the transmission channel. Some trace is invariably left in the channel. We model it as a shift of a local degree of freedom of the channel (the pointer) described by (\ref{a}). }
\label{fig:9}
\end{figure}

Consider first a single-path transmission channel  with a single particle  in the form of a single localized wave packet. The wave packet passes from Alice to Bob, see Fig.~9.
Let us model the interaction of the particle with the transmission channel as von Neumann measurement  with a Gaussian probe. The initial wave function of the pointer is
 \begin{equation}\label{psi0}
 \langle x|\Phi_0\rangle  = \frac{1}{\sqrt{\Delta \sqrt{\pi}}} e^{-\frac{x^2}{2\Delta^2}}.
\end{equation}
When a particle is present in the transmission channel, the interaction  shifts $x$ by $\delta$.
Thus, the state of the measuring device after the interaction is
 \begin{equation}\label{a}
 |\Phi\rangle  =\sqrt{1-\epsilon^2} |\Phi_0\rangle+ \epsilon |\Phi_\perp\rangle,
\end{equation}
where $|\Phi_\perp\rangle$ is orthogonal to the initial pointer state and $ \epsilon=\sqrt{1-e^{-\frac{\delta^2}{\Delta^2}}}$.

How to quantify the strength of the trace? One option is to consider the  probability of detecting  the change in the state of the channel  in an idealized experiment. The probability equals  $\epsilon^2$. Another option is just to use the shift of the pointer via the parameter $\left|\frac{\delta}{\Delta}\right|$.

For a strong trace, the probability criterion does not represent the trace well: it remains almost 1 for $\left|\frac{\delta}{\Delta}\right|=10$ and also for $\left|\frac{\delta}{\Delta}\right|=1000$. In practice, however, it is plausible that in a realistic experiment, when in addition to the quantum uncertainty of the pointer there is an uncertainty of the grid on which we read the pointer, only very large $\left|\frac{\delta}{\Delta}\right|$ can be observed.

For a small value of $\left|\frac{\delta}{\Delta}\right|$, the probability of detection is proportional to the square of this parameter. The transmission of two particles  doubles the shift, but increases the probability of detection by a factor of 4. The linear response seems to be a better representation of the trace.

\begin{figure}
\begin{center}
 \includegraphics[width=7.2cm]{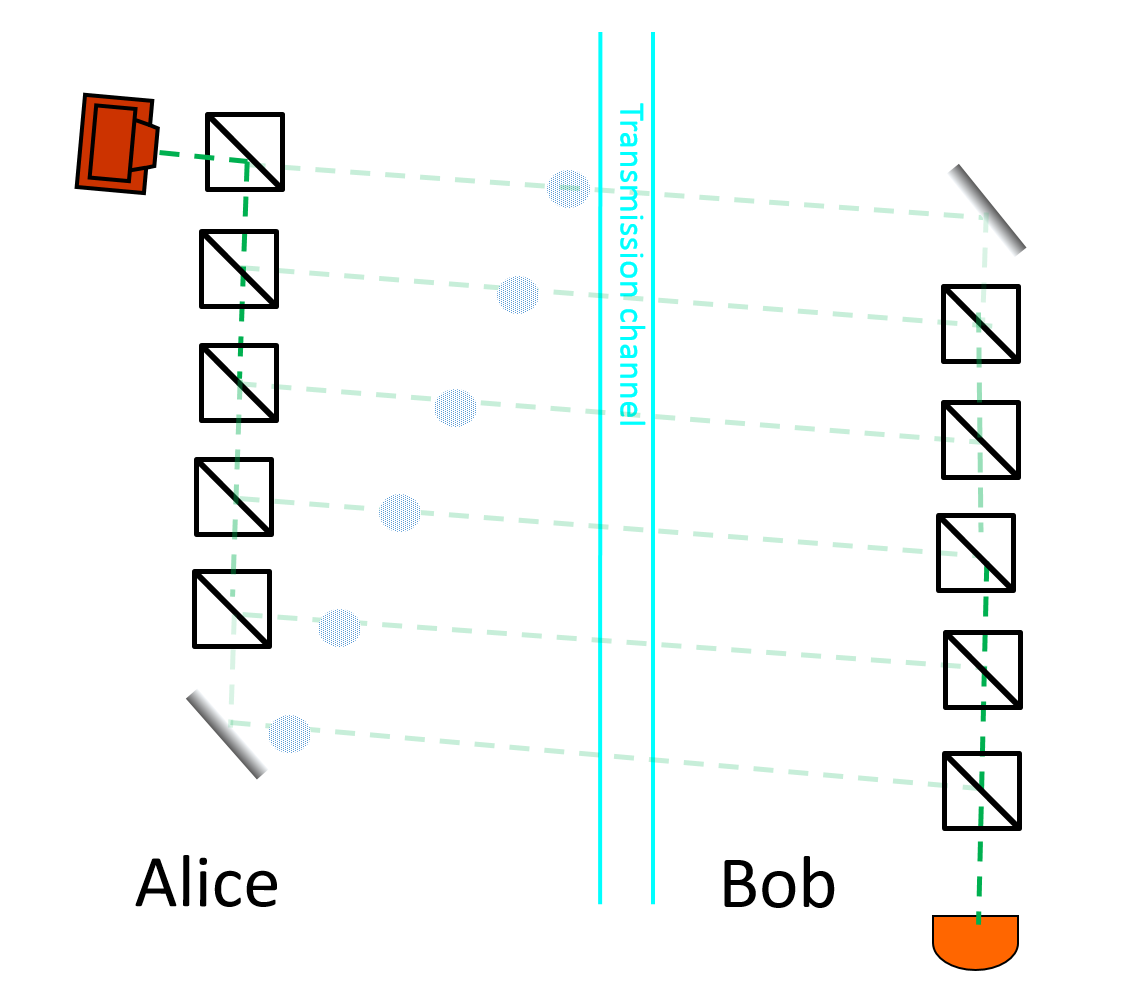}  \end{center}
\caption{ A single particle in a superposition of several localized wave packets passes from Alice to Bob. We assume that the beam splitters are arranged in such a way that all packets have equal amplitudes and the beam splitters on Bob's side are tuned to interfere constructively  toward the detector. }
\label{fig:10}
\end{figure}

 Consider now a single particle passing through the transmission channel which consists of $N$ identical paths as above. The  quantum state of the particle is an equal weight  superposition of  wave packets in all paths, $|\Psi_{in}\rangle=\frac{1}{\sqrt N}\sum_{i=1}^N |i\rangle$,  where $|i\rangle$ signifies the wave packet of the particle inside path $i$ in the transmission channel, see Fig.~10.
 After the particle passes the transmission channel, the state of the particle and the pointers representing the transmission channel  becomes
\begin{equation}\label{1-Nchan}
\frac{1}{\sqrt N}\sum_{i=1}^N
\prod_{j\neq i}| \Phi_0\rangle_j (\sqrt{1-\epsilon^2} |\Phi_0\rangle_i+ \epsilon |\Phi_\perp\rangle_i)|i\rangle.
\end{equation}
  The probability to detect the particle in the transmission channel is the probability to find one of the states $|\Phi_\perp\rangle_i$. It  is  $\epsilon^2$ as in the case of the single-path channel. The sum of the expectation values of the shifts of the $x_i$s    is also the same, $ \sum \langle x_i \rangle = \delta$.

It is important to consider the post-selection measurement of the state of the particle made by Bob. Let  Bob  select the undisturbed state  $|\Psi_{fin}\rangle=\frac{1}{\sqrt N}\sum_{i=1}^N |i\rangle$. This corresponds to detection of the particle by Bob's detector in Fig.~10. For a good, low noise channel, the probability to find this state is very close to 1. The state of the transmission channel then becomes
\begin{equation}\label{1-Nchan}
\frac{N\sqrt{1-\epsilon^2}\prod_{j=1}^N | \Phi_0\rangle_j +\epsilon\sum_{i=1}^N |\Phi_\perp\rangle_i \prod_{j\neq i} |\Phi_0\rangle_j}{\sqrt{ N^2(1-\epsilon^2)+N \epsilon^2}}
.
\end{equation}
At this stage, the probability to find one of the states $|\Phi_\perp\rangle_i$ is reduced dramatically. This is because   the failure of post-selection by Bob implies that the probability to find one of the states $|\Phi_\perp\rangle_i$ is 1. For small $\epsilon$, the probability to detect the particle in the transmission channel after the successful post-selection is approximately $\frac{\epsilon^2}{N}$.

It is interesting that  the post-selection of the particle state does not change the expectation value of the sum of the pointer variables $\langle \sum x_i \rangle = \delta$. One way to see this is to note that $\langle \sum x_i \rangle = \delta$ is proportional to the weak value \cite{AV90} of the sum of the projections on all parts of the channel which equals 1 because the initial state is the eigenstate of the sum of the projections with the eigenvalue 1 \cite{AV91}. For describing the magnitude of the trace, the directions of the shifts are not important. So the relevant parameter is  $\sum_i|\langle  x_i \rangle| $. We have found a lower bound, $\sum_i|\langle  x_i \rangle|\geq \delta $.

 \section{The weak trace }\label{weakpresence}

I analyse next  the trace left in the transmission channel in communication protocols discussed above. All protocols are based on interference, therefore, when they work properly,  only a small trace can be left. I use the same model: in every path of the transmission channel, the presence of the particle  shifts the Gaussian pointer, see  (\ref{a}). I assume that the coupling is weak: $\delta \ll \Delta$, and consequently, $\epsilon \ll 1$. For simplicity, I consider the trace created by particles moving from Alice to Bob and  disregard the trace created on the way from Bob to Alice. Then, the example  described in Fig.~1, is identical to that of  Fig.~9. and the trace in the communication channel is   the shift $\delta$ of the pointer   and the probability to discover the presence of the particle by observing the trace, is $\epsilon^2$. In the communication of the bit 0 using IFM, Fig.~2, and in the protocol with HWPs, Fig.~3 the trace is of the same order of magnitude.

In the IFM communication of the bit 1, see Fig.~2, after the interaction with the transmission channel,
 the state of the particle and the pointer is
\begin{equation}\label{1-IFM}
\frac{1}{\sqrt 2}\left [
  \Phi_0\rangle |L\rangle + (\sqrt{1-\epsilon^2} |\Phi_0\rangle+ \epsilon |\Phi_\perp\rangle)|absorbed\rangle\right].
\end{equation}
If the particle is absorbed by Bob, the trace in the channel is exactly as in the single-path channel of Section \ref{presence}: shift by $\delta$ and the probability to find the trace is $\epsilon^2$. If the particle is detected by Alice, then there is no  trace in the transmission channel. The wave packet ``tagged'' by an orthogonal state of the channel, $|\Phi_\perp\rangle$ cannot reach Alice.

Now I consider the IFM experiment with the chain of the interferometers, Fig.~4, starting with the communication of bit 0, the case in which Bob leaves the interferometers undisturbed. The exact expressions are complicated, but for  weak coupling,   only the first order contibution in $\epsilon$ is significant.
Neglecting the  coupling to the channel, we obtain from (\ref{n-steps}) the state of the particle in the $n$th interferometer
\begin{equation}\label{n steps}
 \cos  \frac{n\pi}{2N}   |L\rangle + \sin \frac{n\pi}{2N}  |R\rangle .
\end{equation}
The wave packet   $|R\rangle$,  ``tagged'' at the $n$th interferometer by the state $|\Phi_\perp\rangle_n$ in the transmission channel,  interferes only with itself and reaches detectors in the state
\begin{equation}\label{N-n steps}
   \cos  \frac{(N-n)\pi}{2N}   |R\rangle -  \sin \frac{(N-n)\pi}{2N}  |L\rangle .
\end{equation}
In this experiment, the particle  ends up in detector $D_2$ (state $|R\rangle$) with the probability close to 1. The state of the transmission channel then is
\begin{equation}\label{afterNchanABC}
{\cal N} \left ( \prod_{n=1}^{N-1} | \Phi_0\rangle_n + \epsilon \sum_{n=1}^{N-1} \sin^2 \frac{n\pi}{2N}\prod_{j\neq n} |\Phi_\perp\rangle_n \right ),
\end{equation}
with normalization   $|{\cal N}|$ close to 1. Therefore, the probability to detect the particle in the transmission channel, i.e., to find at least one of the states $|\Phi_\perp\rangle_n$,
\begin{equation}\label{afterNchanProb}
 \epsilon^2 \sum_{n=1}^{N-1} \sin^4\frac{n\pi}{2N} \sim \epsilon^2 N\frac{2}{\pi}\int_0^{\frac{\pi}{2}} \sin^4x dx = \frac{3\epsilon^2 N}{8},
\end{equation}
 is much larger than the minimal probability to find a single particle  present in this multiple-path channel, which can be as low as $\frac{\epsilon^2}{N}$. Thus, the case of the bit 0 is definitely not a ``counterfactual'' communication. This can also be seen  by calculating the pointers shifts. These shifts are proportional to the expectation value of the projection on the paths of the transmission channel. The shift in path $n$ is $\delta \sin^2 \frac{n\pi}{2N}$ and the sum of all shifts, $\sim\delta   \frac{N}{2}$, is much larger than $\delta$, the standard for the presence of a single particle in the channel.

 The situation is different for communication of bit 1, when Bob blocks the paths of the interferometers, Fig.~4b. Due to Zeno effect, the probability of absorbtion by Bob is negligible. Detector $D_1$ clicks with probability close to 1 telling Alice that the bit value is 1. In this case, there is no trace in the communication channel. It is, therefore,  a counterfactual communication for bit value 1.

Let us turn now to the case of nested interferometers, Fig.~5. The case which is particularly interesting is described in Fig.~5a. Bob does not put the shutter in, and  detector $D_1$ clicks. Alice knows that the bit is 0, and naively, it seems to be a counterfactual communication since the  particle ``could not pass through the transmission channel''. Indeed, the wave packet entering the nested interferometer does not reach detector $D_1$.

The interferometer was  defined only by demanding destructive interferences in particular situations. To make quantitative predictions, we have to specify the beam splitters of the interferometer.  In a possible implementation of the interferometer \cite{Danan},  the first two beam splitters transform the localized wave packet entering the interferometer into a superposition:
\begin{equation}\label{ABC}
|\Psi_{in}\rangle \rightarrow \frac{1}{\sqrt 3}(|A\rangle+|B\rangle+|C\rangle),
\end{equation}
and the other two beam splitters   transform each of the states  as:
\begin{eqnarray}\label{ABC2}
 \nonumber
 |A\rangle &\rightarrow& \frac{1}{\sqrt 3} |1\rangle-\frac{1}{\sqrt 6}|2\rangle +\frac{1}{\sqrt 2}|3\rangle,\\
 |B\rangle &\rightarrow& -\frac{1}{\sqrt 3} |1\rangle+\frac{1}{\sqrt 6}|2\rangle +\frac{1}{\sqrt 2}|3\rangle, \\
 \nonumber
 |C\rangle &\rightarrow& \frac{1}{\sqrt 3} |1\rangle+ \sqrt{\frac{2}{3}}|2\rangle,
 \end{eqnarray}
where states $|i\rangle$ signifies a wave packet entering detector $D_i$, ($i$=1,2,3). It is is easy to see that these rules ensure the required destructive interferences.

 After the interaction, the joint state of the particle and the channel is
\begin{equation}\label{aABC}
  \frac{1}{\sqrt 3} [|A\rangle (\sqrt{1-\epsilon^2} |\Phi_0\rangle+ \epsilon |\Phi_\perp\rangle)+(|B\rangle+|C\rangle)|\Phi_0\rangle].
\end{equation}
From (\ref{ABC2}) we see that the detection of the particle in detector $D_1$  post-selects  the state
$ \frac{1}{\sqrt 3}(|A\rangle-|B\rangle+|C\rangle).$
 Therefore,   the state of the channel after the post-selection is
\begin{equation}\label{a1}
 \sqrt{1-\epsilon^2}|\Phi_0\rangle+ \epsilon |\Phi_\perp\rangle.
\end{equation}
This is exactly the same  state of the channel as in the case that   a single particle passed through it, see (\ref{a}).
 Thus, contrary to the naive expectation, the scheme with nested interferometers does not provide counterfactual communication \cite{Va07}.

\section{The weak trace in ``Direct counterfactual quantum communication''}\label{weakpresenceCount}

Now we are ready to analyze the trace left in the ``Direct counterfactual quantum communication". The case of bit 1, Fig.~6a, is simple. The trace in the communication channel is correlated with the final location of the particle. If it is absorbed by Alice, which happens with a probability close to 1 and corresponds to the proper operation of the protocol, the trace is zero. The wave packets ``tagged'' by  orthogonal states of the channel, $|\Phi_\perp\rangle_{m,n}$, cannot reach Alice.
 The trace is  present only if the particle is absorbed by one of the Bob's shutters which happens with vanishing probability. This is a counterfactual communication protocol for bit 1.

 The more interesting case is that of bit 0, when Bob leaves the interferometer undisturbed, Fig.~6b. We assume that the interaction with the channel is small, $\epsilon \ll \frac{1}{M}$, $\epsilon \ll \frac{1}{N}$, so only the first order in $\epsilon$ should be considered.

 The amplitude of the wave packet of the particle  in the $n$th path of $m$th chain of the inner interferometers $(m,n)$, is
\begin{equation}\label{AMPnm}
  \cos^{(m-1)N} \frac{\pi}{2N} \cos^{(m-1)}  \frac{\pi}{2M}  \sin \frac{\pi}{2M} \sin \frac{n\pi}{2N}.
\end{equation}
A particle present in  the path $(m,n)$  changes  the state of the corresponding pointer according to (\ref{a}):
\begin{equation}\label{M0-step1}
 |\Phi_0\rangle_{m,n} |m,n\rangle \rightarrow \left( \sqrt{1-\epsilon^2} |\Phi_0\rangle_{m,n}+ \epsilon |\Phi_\perp\rangle_{m,n}\right)|m,n\rangle.
\end{equation}
The  wave packet $|m,n\rangle$ ``tagged'' by the orthogonal state $|\Phi_\perp\rangle_{m,n}$ interferes only with itself and leaves the inner chain in the state, see (\ref{N-n steps}):
\begin{equation}\label{N-n stepsSa}
 |m,n\rangle \rightarrow   \sin  \frac{n\pi}{2N}   |R\rangle - \cos \frac{n\pi}{2N}  |L\rangle .
\end{equation}
State $|R\rangle$ is lost and  the state  $|L\rangle$ of the   last inner interferometer of the chain $m$ enters from the right the remaining $M-m-1$  large interferometers. The transformation of this state (which is named $|R\rangle$ in the following equation) in the remaining interferometers is:
\begin{eqnarray}
  \nonumber \label{endlarge}
 |R\rangle&\rightarrow& \sin  \frac{\pi}{2M} \cos^{(M-m-1)N}  \frac{ \pi}{2N} \cos^{(M-m-2)}  \frac{\pi}{2M} \times \\
   &\times& \left ( \cos  \frac{\pi}{2M}   |L\rangle  + \sin \frac{ \pi}{2M}  |R\rangle  \right )+ ...~~,
\end{eqnarray}
where ``...'' signifies the wave packets which do not reach the detectors.

In the protocol, the particle is found with probability close to 1 by detector $D_1$.  After  the detection of the particle, the amplitude of the term $|\Phi_\perp\rangle_{m,n}$   corresponding to detection of the particle in the path $(m,n)$ can be found by collecting the factors in (23-26). It is
\begin{equation}\label{AMPnmPerp}
 \frac{\epsilon}{2} \cos^{(M-2)N} \frac{\pi}{2N} \sin^2 \frac{\pi}{2M} \sin \frac{n\pi}{N} \cos^{(M-2)}  \frac{\pi}{2M}.
\end{equation}
As explained in Section \ref{DirectCF}, the protocol works properly if $1<<M<<N$, so that
\begin{equation}\label{AMPnmappr}
  \cos^{(M-2)N} \frac{\pi}{2N} \sim 1,~~~ \cos^{(M-2)}  \frac{\pi}{2M}\sim 1,
\end{equation}
and the probability that one of the orthogonal states $|\Phi_\perp\rangle_{m,n}$ will be found in the transmission channel is, approximately,
\begin{equation}\label{AMPnmapprox}
\sum_{m,n} \frac{\epsilon^2}{4}   \sin^4 \frac{\pi}{2M} \sin^2 \frac{n\pi}{N} \sim   \frac{\epsilon^2 \pi^ 4 N}{2^7 M^3}.
\end{equation}

The number of paths in the channel is  $\sim MN$. We have seen that a single particle present in such a channel can be found with probability  as low as  $ \frac{\epsilon^2  }{MN}$ which is smaller than the probability of detection of the particle in the protocol by a factor of approximately $ \frac{N^2  }{M^2}$. Since the protocol works well only when $N\gg M$, the trace in the protocol is larger than the trace of a single  particle passing through the channel.

Another criterion of counterfactuality is the sum of displacements of the pointers in all paths of the channel, the standard for which is $\delta$. It can be found by calculating the absolute values of weak values of all projections: $\delta \sum_{m,n}|({\bf P}_{m,n})_w|$ with pre- and post-selection specified by the protocol. The scalar product in the denominator of the weak value is close to 1 since the probability of the post-selection is close to 1. The amplitude of the forward evolving state at path $(m,n)$ is given by (\ref{AMPnm}) and,  similarly, the amplitude of the backward evolving state  is
\begin{equation}\label{AMPnm2}
  \cos^{(M-m-1)N} \frac{\pi}{2N} \cos^{(M-m-1)}  \frac{\pi}{2M} \sin \frac{\pi}{2M} \sin \frac{(N-n)\pi}{2N}.
\end{equation}
Thus, the weak value of the projection on the path $(m,n)$ can be approximated as
\begin{equation}\label{WVProj}
{(\bf P}_{m,n})_w\sim \frac{\pi^2}{8M^2}\sin \frac{n\pi}{N},
\end{equation}
and the sum of all  shifts of  pointers    is
\begin{equation}\label{WVProj}
\delta \sum_{m,n}|({\bf P}_{m,n})_w|  \sim \delta \frac{\pi^2 N}{16M}.
\end{equation}
Since   $N\gg M$, the trace in the protocol is much larger than the standard of the trace of a single particle present in such multiple-path channel.

\section{The weak trace in ``Direct quantum communication with almost invisible photons'' }\label{weakpresence}

Let us turn to the   ``Direct quantum communication with almost invisible photons'' protocol \cite{Li}.
When Bob transmits 0, i.e. does nothing, Fig.~8a, the amplitude in the path $(m,n)$ is
 \begin{eqnarray}\label{Inv1}
 \nonumber
  \sin  \frac{\pi}{2M}  \sin \frac{n\pi}{ N} &{\rm for}& ~m~~ {\rm odd}, \\
  0~~~~~~~~ &{\rm for}& ~ m~~ {\rm even}.
 \end{eqnarray}
The   wave packet $|m,n\rangle$, ``tagged'' by the orthogonal state $|\Phi_\perp\rangle_{m,n}$, interferes only with itself and it leaves the inner chain in the state
\begin{equation}\label{Inv2}
 |m,n\rangle \rightarrow    \sin  \frac{n\pi}{N}   |R\rangle - \cos \frac{n\pi}{N}  |L\rangle .
\end{equation}
The state $|R\rangle$ is lost and  the state  $|L\rangle$ of the   last inner interferometer of the chain $m$ enters from the right the remaining $M-m-1$  large interferometers. Using the fact that  ``tagging'' takes place only for odd $m$,  and that the number of the large interferometers is odd ($M$, the number of beam splitters is even), and that after every second beam splitter the wave function repeats itself,   we can conclude that the wave packet leaves the last beam splitter of the last inner chain  with the same amplitude.  The wave packet entering the last beam splitter of the large interferometers transforms into
 \begin{equation}\label{Inv3}
\cos  \frac{\pi}{2M}   |R\rangle  - \sin \frac{ \pi}{2M}  |L\rangle.
\end{equation}

In the protocol, the particle is detected with  probability close to 1 by detector $D_1$ which detects the state $|L\rangle$.  Thus, after    detection of the particle, the amplitude of the term $|\Phi_\perp\rangle_{m,n}$   corresponding to the detection of the particle in the path $(m,n)$ can be found by collecting factors from (\ref{Inv1}-\ref{Inv3}) and a factor $\epsilon$ due to the interaction:
\begin{equation}\label{AMPnmPerpInv}
 \frac{\epsilon}{2}  \sin \frac{2n\pi}{ N}  \sin^2 \frac{ \pi}{2M}.
\end{equation}
This expression holds only for  odd $m$, the amplitude in the paths with  even $m$ vanishes. Summing the probabilities of finding the record the particle leaves in all the paths, i.e. summing on odd $m$s up to $M-1$ and on integers $n$ up to $N-1$, we obtain the probability of detection of the particle:
\begin{equation}\label{AMPnmapprox}
\sum_{m,n} \frac{\epsilon^2}{4}   \sin^4 \frac{\pi}{2M} \sin^2 \frac{2n\pi}{N} \sim   \frac{\epsilon^2 \pi^4N}{2^8 M^3}.
\end{equation}

A single particle passing through  this transmission channel can be found with probability of   order  $\frac{\epsilon^2}{MN}$. Depending on the ratio $\frac{N}{M}$, it can be  smaller or larger than (\ref{AMPnmapprox}), so we cannot yet decide on the counterfactuality of the protocol.

Compare now the
sum of the pointer shifts in the channel. It can be found by calculating the absolute values of weak values of all projections: $\delta \sum_{m,n}|({\bf P}_{m,n})_w|$ with the pre- and post-selection specified by the protocol. In this case, the  pre- and post-selected states are identical, so the weak values are  expectation values:
 \begin{equation}\label{Inv1111}
({\bf P}_{m,n})_w= \left \{ \begin{array} {cl} \sin^2  \frac{\pi}{2M}  \sin^2 \frac{n\pi}{ N} &{\rm for} ~m~~ {\rm odd}, \\
  0 &{\rm for} ~ m~~ {\rm even}. \end{array} \right.
\end{equation}
and
\begin{equation}\label{WVProj0}
\delta \sum_{m,n}|({\bf P}_{m,n})_w|  \sim \delta \frac{\pi^2 N}{16M}.
\end{equation}

The sum of the pointer shifts when there is one particle in the transmission channel equals $\delta$, so the ratio $\frac{N}{M}$ tells us when  the sum of displacements in the protocol (\ref{WVProj0}) is smaller or larger than that of a single particle present in the channel. We can see  that the criterion of the pointer shifts is in agreement with the criterion of the probability of detection.

Let us now repeat the analysis for the case of bit 1, when Bob puts HWPs in every inner interferometer. Now
 the amplitude in the path $(m,n)$ is
 \begin{eqnarray}\label{Inv4}
 \nonumber
  \sin  \frac{m\pi}{2M}  \sin \frac{\pi}{ N} &{\rm for}& ~n~~ {\rm odd}, \\
  0~~~~~~~~ &{\rm for}& ~ n~~ {\rm even}.
 \end{eqnarray}

The   wave packet $|m,n\rangle$, can be ``tagged'' by the orthogonal state $|\Phi_\perp\rangle_{m,n}$ only for odd $n$. Thus, due to the presence of the HWPs, the wave packet comes back unchanged every second beam splitter. It leaves the chain of the inner interferometers in the state
\begin{equation}\label{Inv5}
 |m,n\rangle \rightarrow    \cos  \frac{ \pi}{N}   |R\rangle - \sin \frac{\pi}{N}  |L\rangle  .
\end{equation}
State $|R\rangle$ is lost, and  the state  $|L\rangle$ of the   last inner interferometer of the chain $m$ enters from the right the remaining $M-m-1$  large interferometers.
 In the chain of the large interferometers it performs usual evolution (\ref{n-steps}) and ends up in the state
\begin{equation}\label{Inv6}
  \sin  \frac{m\pi}{2M}   |R\rangle -\cos \frac{m\pi}{2m}  |L\rangle .
\end{equation}

In the protocol, the particle is detected with  probability close to 1 by detector $D_2$ which detects the state $|R\rangle$.  Thus, after   the detection of the particle, the amplitude of the term $|\Phi_\perp\rangle_{m,n}$   corresponding to the detection of the particle in the path $(m,n)$ can be found by collecting factors from (\ref{Inv4}-\ref{Inv6}) and the factor $\epsilon$ due to the interaction:
\begin{equation}\label{AMPnmPerpInv}
 \frac{\epsilon}{2}  \sin^2 \frac{ \pi}{ N}  \sin^2 \frac{ m\pi}{2M}.
\end{equation}
This expression holds only for odd  $n$, the amplitude in the paths with   even $n$, vanishes. Summing the probabilities of finding the record the particle leaves in all the paths, we obtain the probability of detection of the particle:
\begin{equation}\label{AMPnmapprox2}
\sum_{m,n} \frac{\epsilon^2}{4}   \sin^4 \frac{m\pi}{2M} \sin^4 \frac{\pi}{N} \sim    \frac{\epsilon^2 3\pi^4 M}{2^6 N^3}.
\end{equation}

Again, as in the case of bit 0, the ratio $\frac{N}{M}$ tells us if it is larger or smaller than the minimal probability of detection in case of a single particle present in the transmission channel. However, the dependence is opposite: if for bit 0 the probability of detection in the protocol was smaller than single-particle standard, for bit 1 it will be larger, and vice versa.

The pointer shifts criterion of counterfactuality is in agreement with these results.
The
sum of pointers shifts in all paths of the channel  is proportional to the sum of the weak values of all projections: $\delta \sum_{m,n}|({\bf P}_{m,n})_w|$. Also in this case, the pre- and post-selected states are identical and the weak values are  expectation values:
 \begin{equation}\label{Inv1112}
({\bf P}_{m,n})_w= \left \{ \begin{array} {cl} \sin^2  \frac{m\pi}{2M}  \sin^2 \frac{\pi}{ N} &{\rm for} ~n~~ {\rm odd}, \\
  0 &{\rm for} ~ n~~ {\rm even}, \end{array} \right.
\end{equation}
and
\begin{equation}\label{WVProj1}
\delta \sum_{m,n}|({\bf P}_{m,n})_w|  \sim \delta \frac{\pi^2 M}{4N}.
\end{equation}
The ratio $\frac{N}{M}$ tells us when  the sum of displacements in the protocol (\ref{WVProj1}) is smaller or larger than that of a single particle. If it is smaller for bit 1, it is larger for bit 0, and vice versa.

\section{Security of Counterfactual Protocols}\label{secur}

One of the motivations for ``counterfactual'' protocols, in which no particles are present in the transmission channel, is that it is secure against Eve who is trying to eavesdrop the communication: she has no particles to look at \cite{CFseq}. The obvious cryptographic weakness of the protocols with shutters which are 100\% counterfactual is that Eve can use an active attack detecting the presence of the shutters. Moreover, the shutter can be detected by Eve in a counterfactual way \cite{CFattack}, and recently there has been a claim of a very efficient attack of this kind  \cite{CFInt5}.

If Eve uses an active attack, probing  Bob's bit by sending her particles, the counterfactuality property does not help. So, for the analysis of counterfactuality we should only consider passive attacks in which Eve ``eavesdrops'', i.e. measures the presence of the particle in the transmission channel. Under this condition, the counterfactual protocols with shutters are secure.

Consider the following counterfactual key distribution protocol. There are two identical chains of interferometers as in  Fig.~4. One of the chains is defined as bit 0, and  the other as bit 1. On each run of the protocol, Alice randomly sends a single particle through one of the interferometers and Bob randomly chooses one of the interferometers and places the shutters in all of its paths. Every time Alice detects the particle in detector $D_1$ of one of the interferometers, she makes a public announcement. Detector $D_1$ can click only if Alice and Bob chose the same interferometer, i.e. they chose the same bit. This creates a common key.

 The multiple shutters Bob placed represent the weak point of the protocol due to the reason above, although we can improve it using detectors instead of shutters and telling Alice to send particles not every time, but only at some random times (using ideas of \cite{GV}). Anyway, we made a postulate here that Eve only performs some (weak) nondemolition measurements of the presence of the particle running in the interferometer. Let us see if Eve can get some information about the key in this way.

 If Eve detects the particle in one of the interferometers, it cannot be the one which generates a correct bit in the key. For a correct bit generation event, Alice and Bob have to choose the same interferometer.   Eve can detect the particle only if it is present, i.e., it has to be chosen by Alice. If Bob also chooses  this interferometer, then after detection by Eve, the particle  has to be absorbed by Bob, so it will not reach Alice.

 Only if the interferometer is not chosen by Bob, the particle seen by Eve in the nondemolition measurement can reach Alice, and there is a nonzero probability that the detector $D_1$ will click and thus Alice will declare generation of a bit in the key. But this will be an error bit, since Bob has chosen the other interferometer. Eve introduces errors, and the only information she gets is about these error bits.

 Let us turn to the ``Direct counterfactual quantum communication'' protocol. When Bob places the shutters, Fig.~6a, Eve cannot get information about the correct bit because Eve's detection causes the loss of the particle. (It is not surprising, since in this case the protocol is counterfactual.)

 If the bit is 0, and the interferometer is free, Fig.~6b, Eve's detection will not necessarily  lead to the loss of the particle. Detection of the particle by Eve in the last chain of inner interferometers will lead to part of its wave packet to enter the last beam splitter from the right, so  most probably it will create an error: $D_2$ clicks, but Bob's bit is 0. However, Eve's detection of the particle in any of the first $M-2$ chains of inner interferometers will lead to part of the particle wave packet to enter the last beam splitter from the left side, so most probably   $D_1$ will click, corresponding to a successful transmission of the correct bit.   Thus, sometimes, Eve  gets a reliable information about the transmitted bit.

 Eve, who eavesdrops only by observing the original particle of the communication protocol, cannot learn any correct bit of real counterfactual protocols with shutters which transmit bit 1, but she does learn some correct bits in ``counterfactual'' communication protocols of bit 0. This provides another argument why such  protocols should not be named counterfactual.

\section{ Counterfactuality according to the Bohmian Mechanics}\label{Bohm}

It seems that Bohmian mechanics \cite{Bohm52}, which postulates that every particle has a trajectory, should provide the ultimate answer regarding the counterfactuality of a protocol. There are   unambiguous answers to criteria (2) and (3) of Sec. \ref{Criteria}: either the Bohmian trajectory passes through the transmission channel, or it does not.
However, the fact that the Bohmian trajectory does not pass through the transmission channel does
not tell us that Eve, who has an access to this channel, cannot get some information about this communication.

For completeness of the counterfactuality  analysis, I will consider  the following technical question.  Is  the Bohmian trajectory    present in the transmission channel of the protocols discussed in this paper?

Bohmian position of a particle cannot be in a place where the (forward evolving) wave function vanishes. Therefore, the successful IFMs of the presence of an opaque object, described in Fig.~2 and  Fig.~4b, are counterfactual according to the Bohmian trajectory criterion. Moreover, the IFM of the absence of an opaque object,  Fig.~5a, is counterfactual too. Direct counterfactual communication \cite{Salih}, as its predecessor \cite{Ho06} and their variation presented in Fig.~6,  are counterfactual. In all these protocols there is no continuous path with non-vanishing wave function between the source and the detector the particle reached.

\begin{figure}
\begin{center}
 \includegraphics[width=7.2cm]{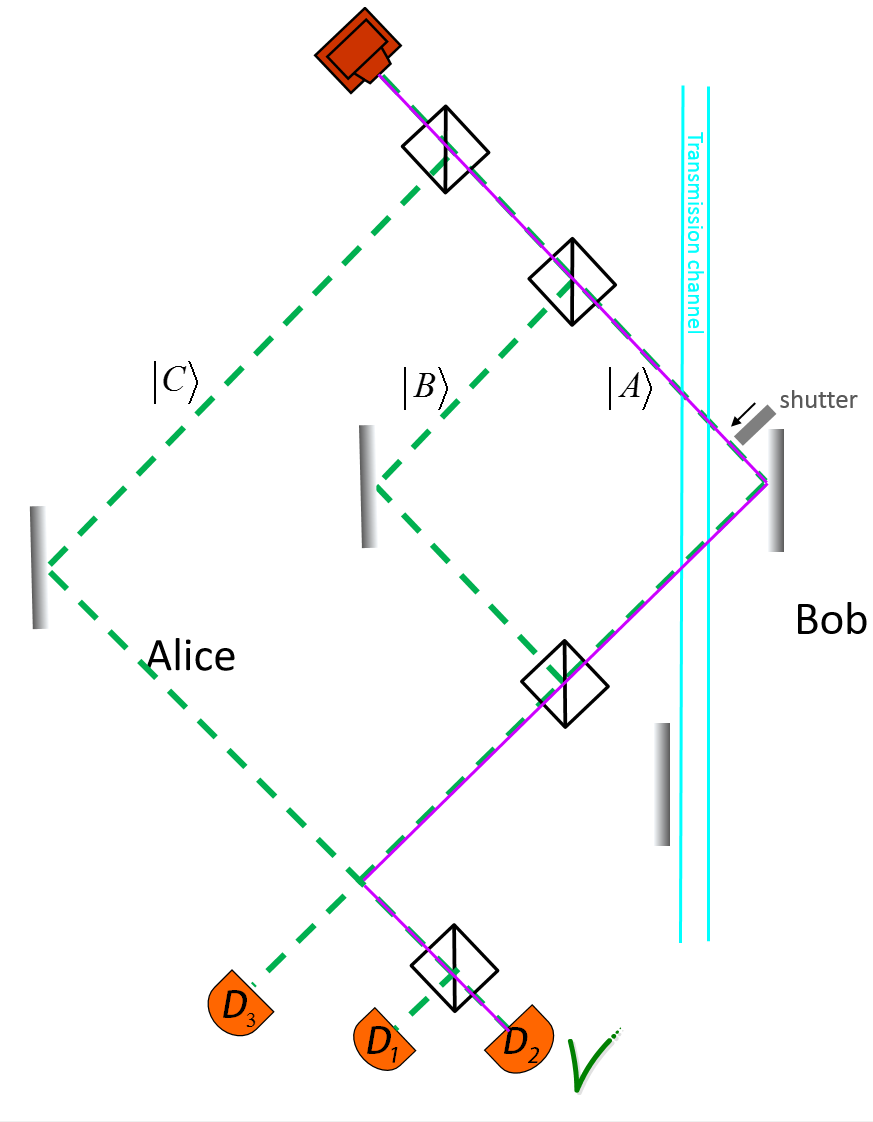} \end{center}
\caption{ The modification of the experiment shown on Fig.~5a which is not counterfactual according to the Bohmian criterion. The interferometer is tuned such that there is a destructive interference toward $D_2$ when the shutter is present. Thus, if detector $D_2$ clicks, we know that  path $A$ is free. Although naively, the particle reaching $D_2$ cannot pass through the inner interfereometer, the Bohmian trajectory  of the particle (solid line) does pass through $A$. }
\label{fig:11}
\end{figure}

In the ``Direct quantum communication with almost invisible photons'' \cite{Li}, see Fig.~8,  the probability that the Bohmian particle will pass through the transmission channel can be found from the maximal amplitude in the paths passing through the channel.  For bit 0, the amplitude is given by (\ref{Inv1}), and therefore, the maximum probability is approximately $\frac{\pi^2}{4M^2}$.  For bit 1, the amplitude is given by (\ref{Inv4}), and the maximum probability is approximately $\frac{\pi^2}{N^2}$. Thus, for  large $M$ and $N$, the probability that the communication is counterfactual is close to 1 for both bit values.

However, it should be mentioned, that in most cases when the communication is not counterfactual,  the Bohmian particle  crosses the transmission channel, not once, but   many times. Given equal probability for bit values, the expectation value of the number of crosses of the transmission channel by the Bohmian particle can be obtained from the sum of the probabilities on all paths. The amplitudes are given by (\ref{Inv1}) and (\ref{Inv4}). Consequently, the expectation value of the number of crosses is, approximately, $\frac{\pi^2}{4}\left (\frac{M}{N}+\frac{N}{4M}\right)$, and it cannot be much less than 1 for any choice of  $M$ and $N$.

Note that in  the ``Direct counterfactual communication protocol''  {\it experiment} \cite{Salih}, the expectation value of the number of crosses of the transmission channel is also   of   order  1, but all these events  happen when the particle is lost, and these cases are discarded according to the protocol.

\begin{figure}
\begin{center}
 \includegraphics[width=7.2cm]{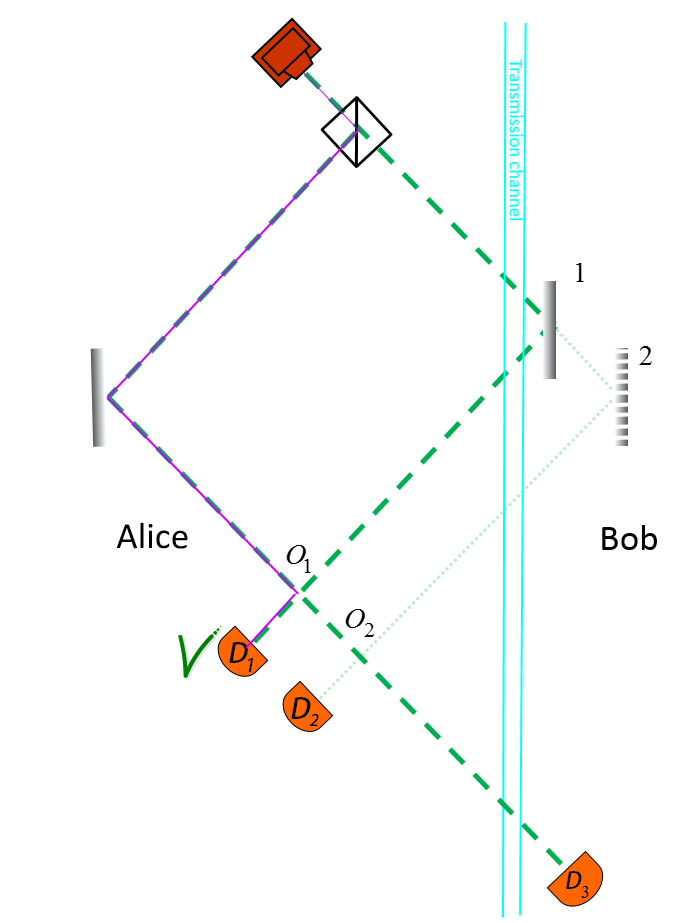} \end{center}
\caption{ Simple protocol which is counterfactual according to Bohmian criterion. Bob places the mirror in the position 1 or 2. This makes clicks of Alice's corresponding detectors $D_1$ or  $D_2$ possible. Only empty wave passes through the transmission channel when Alice's detector clicks. }
\label{fig:11}
\end{figure}

I mentioned above that the protocol presented in Fig.~5a, which  demonstrates  the absence of an object in the apparently interaction-free manner,  is counterfactual according to Bohm. It is interesting that a slight modification of this setup, similar to the experiment actually performed \cite{Danan},  is {\it not} counterfactual. Consider a setup presented in Fig.~11. A straightforward calculations of Bohmian trajectories (similar to that of Bell \cite{Bell}, or in a simplified way \cite{gili}) show that {\it all} particles detected by $D_2$   passed through arm $A$.

In fact, the experiment \cite{Danan} indicated the presence of the weak trace  in $A$, but it had nothing to do with the presence of the Bohmian trajectory there. The weak trace appeared also in $B$ and $C$ where the Bohmian particle was not present.

Empty waves cause observable difference when they ``catch'' Bohmian particles. This allows a very simple communication protocol which is counterfactual  according to Bohmian criterion, see Fig.~12. Particles are sent through a MZI without a second beam splitter at particular times.  One of the mirrors is in  Bob's place and he  has two options for placing it, such that the wave bouncing off the mirror will end up in detector $D_1$ or $D_2$ on Alice's side. The wave packet going through another arm ends at detector $D_3$ on Bob's side. If the particle does not reach $D_3$, Bob knows that it reaches Alice and that {\ it he} has chosen the detector which will detect the particle. Every time $D_1$ or $D_2$ clicks the Bohmian particle does not pass through Bob's place. Only an empty wave was directed by Bob's mirror. Indeed, the wave packets overlap at point $O_1$ or $O_2$  and the Bohmian particle must  ``change hands'' in the overlap, so Bohmian particles reaching Alice's detectors never cross communication channel.

Probably the majority would not consider the communication protocol described in Fig.~12 as counterfactual. According to Wheeler's common sense argument \cite{Whe}, the particle reaching Alice could come only through Bob's site. This  example, and Englert et al. setup \cite{SUR}, in which a strong trace (observed at a later time) is left in a place where the Bohmian particle was not present, explain  why  the Bohmian criterion of counterfactuality does not agree with the intuition of most physicists.

\section{Conclusions }\label{conc}

The standard quantum formalism, in contrast to Bohmian mechanics, does not specify the position of a quantum particle. Thus, it does not provide an unambiguous answer to the question: Is a particular communication protocol counterfactual? I.e.: Was the particle present in the transmission channel? In this paper I analyzed an approach to answering this question based on the weak trace the particle leaves in the channel. I compared the trace left in the channel in  recently proposed protocols claimed to be counterfactual with the trace in the protocol constructed to transmit a single particle in the same channel.

In the analysis, I considered two criteria for comparing  the traces. First, the probability of finding a conclusive evidence for the presence of the particle, and second, the expectation value of the sum of total shifts of some variables of the channel.

The question of counterfactuality of the protocols is considered in cases the protocols work properly, i.e. when the particle is detected by the right detector. It means that the particle is pre- and post-selected. The protocols were compared with the transmission of a single pre- and post-selected particle. In all these cases the probability of post-selection was closed to 1.

The analyses using the two criteria of the trace led to the same conclusion. {\it It is possible to communicate only one value of a bit in a counterfactual way.}

 The protocol  ``Direct counterfactual quantum communication" \cite{Salih} is fully counterfactual for bit value 1. The trace is identically 0. Nothing changes in the transmission channel and therefore there is zero probability to detect the particle in the transmission channel. Passive eavesdropping provides no information about the transmitted bit.

 However, the protocol is not counterfactual for the bit value 0. It is true that by increasing the number of paths in the channel, the probability of finding a conclusive evidence of the presence of the particle reduces, but increasing the number of paths also reduces the probability to find a single quantum particle when it passes the channel.  The   probability to find the presence of the particle in the transmission channel in the event of successful operation of the protocol is larger than the probability to detect a particle successfully passing through this channel. Eve, using passive attack, obtains some information about the transmitted bit.

 The criterion of the  shifts of variables of the channel tells us the same. The expectation value of the sum of the shifts is zero for bit 1, but for bit value 0  it is larger than the sum of the shifts when a single particle passes the channel.

 The protocol  ``Direct quantum communication with almost invisible photons'' is not fully counterfactual for any bit value. Some trace is always left in the transmission channel. However, if we are ready to consider a protocol as counterfactual when it leaves a trace which is much smaller than the trace of a single particle passing through this channel, then we can arrange that it will be counterfactual for one of the bit values. By playing with the numbers $N$ and $M$ of the inner and the external interferometers respectively, the protocol can be made counterfactual for value 0 or for value 1 of the bit. It {\it cannot} be made counterfactual for both.

 The analysis of the trace left by a particle passing through a $N$-path channel of Section VI showed a surprising result:  the probability of detection in the channel of the successfully transmitted particle  is reduced by the factor of   $\frac{1}{N}$. It helped me to analyse the counterfactuality of the protocols, but it might also open new avenues   for useful  quantum communication  applications.

 I also hope that this study will  lead to a deeper understanding of the question: ``Where are particles passing through interferometers?'' \cite{past,morepast,Bart,Jordan,Poto}.

I thank Yakir Aharonov, Eliahu Cohen and Shmuel Nussinov for helpful discussions. This work has been supported in part by the  Israel Science Foundation  Grant No. 1311/14  and the German-Israeli Foundation  Grant No. I-1275-303.14.

\end{document}